\documentclass{elsart}

\usepackage{graphicx}
\usepackage{amssymb,amsmath}

\newcommand {\kB}{%
  \ensuremath{k_{\mathrm{B}}}}
\newcommand {\Tc}{%
  \ensuremath{T_\mathrm{c}}}
\newcommand {\Ic}{%
  \ensuremath{I_\mathrm{c}}}
\newcommand {\TBKT}{%
  \ensuremath{T_{\mathrm{BKT}}}}
\newcommand {\vs}{%
  \ensuremath{v_\mathrm{s}}}
\newcommand {\ns}{%
  \ensuremath{n_\mathrm{s}}}
\newcommand {\Ns}{%
  \ensuremath{N_\mathrm{s}}}
\newcommand {\xiGL}{%
  \ensuremath{\xi_{\mathrm{GL}}}}
\begin{document}

\begin{frontmatter}

\title{Fluctuation effects in superconducting nanostrips}

\author[unizh,DLR]{A. Engel\thanksref{author}}
\author[DLR]{A.D. Semenov}
\author[DLR]{H.-W. H\"ubers}
\author[unik]{K. Il'in}
\author[unik]{M. Siegel}
\address[unizh]{Physics Institute of the University of Zurich, Winterthurerstr.\ 190, 8057 Zurich, Switzerland}
\address[DLR]{DLR Institute of Planetary Research, Rutherfordstr.\ 2, 12489 Berlin, Germany}
\address[unik]{Institute of Micro- and Nano-Electronic Systems, University of Karlsruhe, Hertzstr.\ 16, 76187 Karlsruhe, Germany}
\thanks[author]{Corresponding author. Tel: +41-1-635-5776; fax: +41-1-635-5704; e-mail address: andreas.engel@physik.unizh.ch}

\begin{abstract}
Superconducting fluctuations in long and narrow strips made from ultrathin NbN films, have been investigated. For large bias currents close to the critical current fluctuations led to localized, temporary transitions into the normal conducting state, which were detected as voltage transients developing between the strip ends. We present models based on fluctuations in the Cooper pair density and current-assisted thermal-unbinding of vortex-antivortex pairs, which explain the current and temperature dependence of the experimental fluctuation rates.
\end{abstract}

\end{frontmatter}

% main text
\section{Introduction} \label{intro}

Over the last decade a whole range of superconducting quantum detectors \cite{Peacock96} and elements suitable for quantum computing \cite{Nakamura99} have been proposed. These novel devices exploit various aspects of the superconducting state and most of them are based on superconducting thin films or even micro- and nanostructures. For all of these devices to offer superior advantages compared to more conventional semiconducting devices low noise levels are one of several prerequisites. For macroscopic superconductors noise levels are usually very low because of the low operating temperatures and the formation of the superconducting energy gap $\Delta(T)$. However, as the size of the superconductor becomes comparable to the superconducting coherence length $\xi(T)$ in one or more dimensions fluctuation effects play an increasingly important role \cite{Skocpol75} in the overall performance of the device.

In this letter we will discuss fluctuation effects in superconducting nano-strips near the $I$-$T$ phase transition. In previous studies fluctuations in superconductors with reduced dimensions have often been done in the limit of small applied currents \cite{Tinkham96%,Wilson01
}. In contrast, we consider fluctuations in the opposite limit of very high applied currents \cite{Knoedler82} close to the critical current $I_c(T)$, a situation that is often encountered in real applications, e.g.\ transition edge sensors \cite{Lee96}. We have realized an experimental setup, where superconducting fluctuations can be detected as ultra-fast voltage transients developing between the ends of a long and narrow superconducting strip. The detected voltage pulses had typical durations of 10~ns or less. Due to the high applied current even small deviations from the equilibrium state, such as e.g.\ fluctuations or the absorption of visible or near-infrared photons, led to a local and temporary transition into the normal conducting state. The number of voltage transients per second is defined as the voltage pulse rate $\Gamma$ and was measured as a function of temperature and bias current. For the analysis of the experimental results we will consider two independent thermodynamically driven fluctuation types: (i) fluctuations of the number of quasi-particles (QP). Macroscopically this may be expressed as localized fluctuations in the critical current density. If the critical current density drops below the applied current density, the device will switch into a temporary resistive state. (ii) fluctuations of temporarily created bound vortex-antivortex pairs (VAPs). Under the acting Lorentz force due to the relatively high applied current density such bound vortices may unbind and move towards opposite edges of the strip \cite{Knoedler82}. Such vortex motion also leads to a resistive state.

\section{Fluctuations in the number density of QP} \label{sec.pairflucs}
\begin{figure}
\begin{center}
\includegraphics[width=0.8\textwidth]{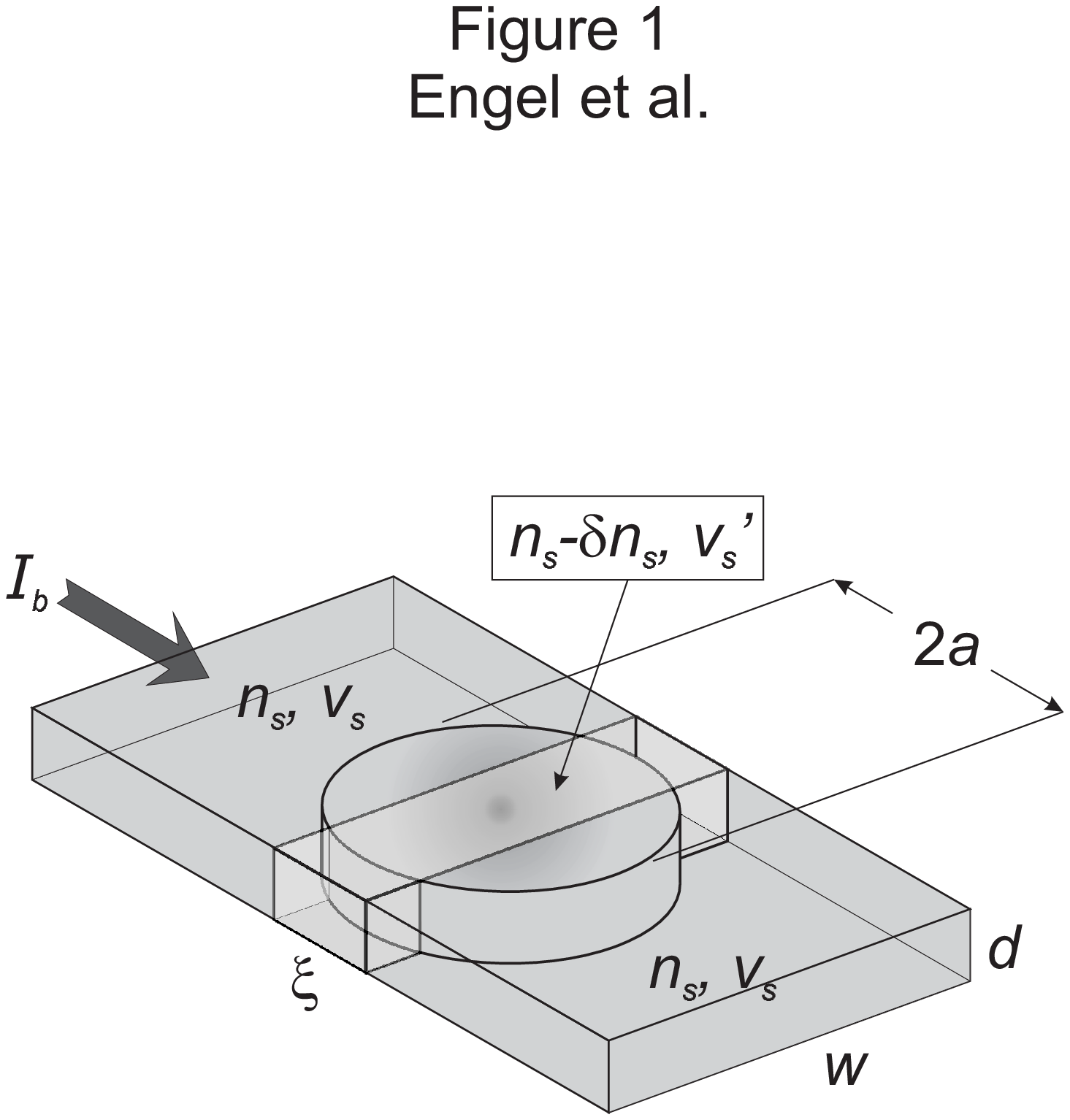}
\end{center}
\caption{Schematic drawing of the refined hotspot model. The grey shaded volume is a cloud with reduced
Cooper-pair density $\ns-\delta\!\ns$. If the density is low enough such the mean pair velocity $\vs'$
exceeds the critical velocity $\vs^\ast$ in a cross-section with longitudinal extension $\xiGL$ or larger, this
cross-section switches into the normal conducting state. \label{fig.schematic}}
\end{figure}

The number density of QP in superconductors are subject to thermally activated fluctuations \cite{Wilson01,Wilson04}. To evaluate their role in the voltage pulse rate in these superconducting strips we make use of a recently proposed refinement of the hotspot model in superconducting single photon detectors \cite{Semenov04}. Suppose in a disc-like volume $V$ with thickness equal to the film thickness $d<\xiGL(T)$, the Ginzburg-Landau (GL) coherence length\footnote{The GL coherence length \xiGL\ is related to the BCS coherence length $\xi_0$ in dirty superconductors; $\xiGL=\sqrt{\xi_0 l}$, with $l$ the mean free path of the charge carriers.}, and radius $a$, $\xiGL(T) \leq 2a \leq w$, the number of Cooper pairs \Ns\ decreases by $\delta\!\Ns$, see also Fig.~\ref{fig.schematic}. Far from the fluctuation site and at temperatures well below the critical temperature \Tc\ the current density is proportional to the velocity of the superconducting charge carriers \cite{Tinkham96}, thus the current may be expressed as $I=\ns \overline{\vs}\,2e\,wd$, with \ns\ the number of Cooper pairs per unit volume, $\overline{\vs}$ the mean velocity of Cooper pairs and $2e$ their charge. Furthermore, we assume the current density to be homogeneous over the cross-section of the strip, because of the effective magnetic penetration length $\Lambda=\lambda_{\mathrm{L}}(0)^2/d$ being much larger than the strip width and thickness. For a cross-section including the fluctuation volume the Cooper pair density is not homogeneous, anymore. However, one can easily show within GL theory that the supercurrent redistributes to keep the mean velocity of the superconducting charge carriers the same all over the cross-section as long as there is no magnetic flux linked to the fluctuation. Thus, the current is now given by $I=n_{\mathrm{s,eff}} \overline{\vs'}\,2e\,wd$, with $n_{\mathrm{s,eff}}=\ns - \delta\!\Ns/V$ and $\overline{\vs'}>\overline{\vs}$. The minimum critical fluctuation that drives the cross-section normal conducting is then given by the condition that $\overline{\vs'}$ is equal to the critical velocity $\overline{\vs}\,^\ast \approx \Ic/(\ns\,wd)$, \Ic\ is the temperature dependent depairing critical current. Using simple algebra this can be converted into the minimum number of QP that need to be destroyed within the fluctuation volume
\begin{eqnarray}
\delta\!N^\ast & = & \pi\,awd\,\ns\left(1-\frac{I}{\Ic}\right)\label{eq.1}\\
                 & \approx & \frac{\pi}{2}\,awd\,N_0 \Delta(I,T)\left(1-\frac{I}{\Ic}\right)\label{deltaN},
\end{eqnarray}
where in Eq.\ (\ref{deltaN}) we have used $\ns \approx (N_0/2)\Delta(I,T)$ for temperatures $T \ll \Tc$, with $N_0$ the density of states at the Fermi energy and $\Delta(I,T)$ the superconducting energy gap or pairing potential. In above expressions we use the depairing critical current. Whether the experimentally determined critical current value is indeed the depairing critical current or rather the depinning critical current for vortex motion is not easily determined. This issue is further discussed in Sec.\ \ref{sec.vap}.

The thermodynamic probability for such a fluctuation is given by $\exp[-\Delta F/\kB T]$ using the change in free energy $\Delta F$ associated with the fluctuation, \kB\ is the Boltzmann constant. For the change in free energy we used the minimum excitation energy $E_{\mathrm{min}}(I,T)$ required to break one Cooper pair, i.e.\ create two additional QP, times the number new QP:
\begin{equation}
\Delta F = \frac{\delta\!N}{2} E_{\mathrm{min}}(I,T)\label{deltaF}
\end{equation}
with the minimum excitation energy given as
\begin{eqnarray}
E_{\mathrm{min}}(I,T) & = & \Delta(I,T)\left(1-\frac{I}{\Ic}\right) \\\nonumber
       & = & \Delta_0 \left[1-\left(\frac{T}{\Tc}\right)^4\right]^{\frac{1}{2}} \left[1-\alpha
             \left(\frac{I}{\Ic}\right)^2\right] \left(1-\frac{I}{\Ic}\right).\label{excE}
\end{eqnarray}
In the last expression $\Delta_0=1.76 \kB \Tc$ is the BCS energy gap at zero temperature and the following term is a relatively simple analytical approximation to the BCS temperature dependence of the energy gap. An applied bias current further reduces the energy gap \cite{Levine65,Anthore03}. This contribution is generally of minor importance except at temperatures very close to \Tc. We account for this contribution by the phenomenological parameter $\alpha = 0.03 (1-T/\Tc)^{-0.5}$. The dominant current-induced effect is a shift of the Fermi sphere in $k$-space which reduces the minimum excitation energy $E_{\mathrm{min}}=\Delta - p_\mathrm{F}\vs$, $p_\mathrm{F}$ is the momentum corresponding to the Fermi-energy \cite{Tinkham96}. Furthermore $\Delta \approx p_\mathrm{F}\vs^\ast$ and using the above mentioned proportionality between the superconducting current and the speed of its carriers, we arrive at $E_\mathrm{min} \propto (1-I/\Ic)$.

Experimentally measured fluctuation rates are not only caused by minimum fluctuations needed to trigger a voltage pulse but by all fluctuations temporarily leading to more QP than $\delta\!N^\ast$. Although larger fluctuations require more energy and are less frequent we take the integral over thermodynamic probabilities for fluctuations from the minimum number $\delta\!N^\ast$ to the maximum number $2\ns\,\pi a^2 d$. Furthermore, one has to take a second integral over the fluctuation volume given by the radius $a$. The minimum volume is given when all Cooper pairs within that volume have to be broken in order to cause a voltage pulse. It follows that the minimum radius is given by $a_{\mathrm{min}}=(1-I/\Ic)w/2$. If the bias current is close enough to \Ic\ the minimum radius $a_{\mathrm{min}}$ becomes less than $\xiGL(T)$. Because variations on a scale smaller than $\xiGL(T)$ do not influence superconductivity an absolute minimum of $a_{\mathrm{min}}=\xiGL(T)/2$ was set. As the upper limit of the fluctuation volume we defined the strip width $w$. Then, the overall thermodynamic probability for a dark count event caused by fluctuations of the QP number density results as
\begin{equation}
P(I,T) = \int_{a_{\mathrm{min}}}^{w/2} \int_{\delta\!N^\ast}^{2\ns\,\pi a^2 d} \frac{Lw}{a^2}
         \exp\left(-\frac{\Delta F}{\kB T}\right) \d \delta\!N \d a, \label{prob1}
\end{equation}
with
\begin{equation}
a_{\mathrm{min}} = \begin{cases}
                   (1-I/\Ic)\frac{w}{2} & \mathrm{for}\ (1-I/\Ic)w \geq \xiGL(T)\\
                   \frac{\xiGL(T)}{2}     & \mathrm{else.}
                   \end{cases} \label{mina}
\end{equation}
Fluctuations of different size $a$ are weighted with the number of independent fluctuation sites $Lw/a^2$. The fluctuation rate is obtained by multiplying the probability of Eq.\ (\ref{prob1}) with an attempt rate $\Omega_{0,\mathrm{cp}}$. In general, the attempt rate also depends on the the bias current, temperature and size of the fluctuations. Lacking a well-founded theoretical expression for the attempt rate we use it as an adjustable parameter to fit the experimental data.

\section{Thermal unbinding of vortex-antivortex pairs} \label{sec.vap}
Another type of fluctuations is intimately linked to the Berezinskii--Kosterlitz--Thouless (BKT) transition in thin superconducting films \cite{Berezinskii70,Berezinskii71,Kosterlitz73}. Below the BCS transition temperature \Tc\ in zero magnetic field excitations of the form of bound VAPs with no net flux may exist in two-dimensional films. Between the BKT transition temperature \TBKT\ and \Tc\ bound pairs and thermally activated unbound vortices coexist. In the absence of pinning the unbound vortices lead to a finite resistance even below \Tc. If a transport current is applied it exerts a Lorentz force which is opposite in direction for the two partners of a VAP. Thus no net force acts on a bound pair, but it leads to a preferred orientation of the pair and a reduction of the binding energy. Consequently, the combined action of the applied current and thermal activation may lead to pair-unbinding even below \TBKT. This current-assisted thermal unbinding (CATU) of VAPs may in turn be an important source of voltage pulses in the superconducting strips under investigation and is explored in more detail below.

\begin{figure}
\begin{center}
 \includegraphics[width=0.8\textwidth]{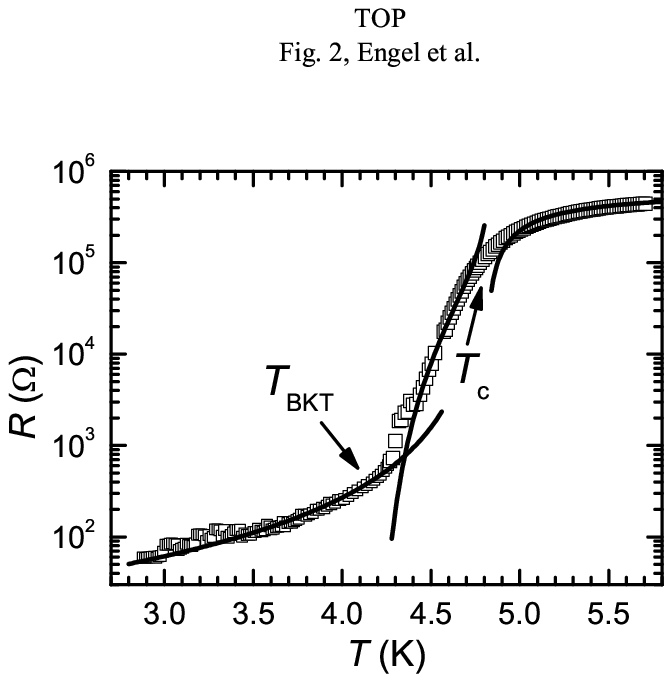}%
\end{center}
 \caption{Resistive transition of a 5 nm thick NbN meander. The solid lines are least-square
 fits for the various temperature intervals as described in the text. The BCS and BKT transition
 temperatures are indicated by arrows.\label{fig.RvsT}}
\end{figure}
Whether or not the above sketched BKT transition is an appropriate model for such NbN strips one should make sure that the strips fall within the two-dimensional (2D) limit and vortices can form in the strips. The NbN films have typical GL coherence lengths extrapolated to zero temperature of approximately 4 to 5~nm. With minimum strip widths of about 80 nm, $\xiGL(T)<w$ for all temperatures except very close to \Tc. Likharev \cite{Likharev79} gives a criterium for the minimum strip width for the existence of vortices: $w \geq 4.4\,\xiGL(T)$. This more stringent criterium is also fulfilled for the temperatures of interest, i.e.\ for which fluctuation rates were measured. These theoretical considerations are backed by an analysis of the resistive transition of a sample shown in Fig.\ \ref{fig.RvsT} together with appropriate least-square fits for the different temperature intervals (solid lines). The resistance data was collected with a standard 4-wire setup and the sample immersed in liquid or gaseous helium at temperatures below and above 4.2 K, respectively. The rounding of the transition above \Tc\ is assumed to be due to fluctuation conductivity of the Aslamazov-Larkin type \cite{Aslamazov68} in a dirty superconducting 2D-film with a high sheet resistance
\begin{equation}
\sigma_{\mathrm{2D}}=\frac{1}{16}\frac{e^2}{\hbar d}\frac{\Tc}{T-\Tc} \label{ALcond},
\end{equation}
with $e$ the elementary charge and $\hbar=h/2\pi$ Planck's constant. Fitting of the resistance data in the relevant temperature region using Eq.\ (\ref{ALcond}) allowed to extract the transition temperature $\Tc=4.81$ K and the normal state sheet resistance $R_\mathrm{N}=1.4$ k$\Omega$. Below \Tc\ the resistance is dominated by thermally unbound VAPs and the BKT theory gives its temperature dependence as \cite{Mooij84}
\begin{equation}
R \propto \exp\left[-2\left(b\,\frac{\Tc-T}{T-\TBKT}\right)^{1/2}\right] \label{RBKT},
\end{equation}
with $b$ a parameter from the theory. For an infinite 2D-film the resistance should vanish below the BKT transition temperature, which turned out to be $\TBKT=4.14$ K for this meander. In a finite sample, however, free vortices are present even below \TBKT\ and Mooij \cite{Mooij84} gives the following relation for the sheet resistance:
\begin{equation}
\frac{R_\mathrm{s}}{R_\mathrm{N}} = 2\pi\,y(l_w)\,\exp(-2l_w), \label{Rfinite}
\end{equation}
with a length scale $l_w=\ln[w/\xiGL(T)]$ and $y(l_w)$ is the pair excitation probability containing the temperature dependence. Following some of the simplifications in Ref.\ \cite{Mooij84} Eq.\ (\ref{Rfinite}) can be expressed as
\begin{equation}
\frac{R_\mathrm{s}}{R_\mathrm{N}} \approx \left(\frac{\xiGL(T)}{w}\right)^\eta
\end{equation}
and the exponent $\eta$ is of the order of 1. Using the simple GL-relation $\xiGL(T)=\xiGL(0)(1-T/\Tc)^{-1/2}$ with $\xiGL(0)=4.2$~nm, and the other parameters as already determined, we can fit the temperature dependence in this regime with $\eta=3.7$. The above sketched analysis gives a self-consistent description of the resistive transition in terms of a BKT-transition in a finite 2D superconducting film. However, there is a small temperature range of about $0.2$~K just below \Tc, where the Likharev criterium prohibits the formation of vortices due to the large value of $\xiGL(T)$. In this region alternative explanations based on e.g.\ Josephson-junction arrays \cite{Johnson98} might be necessary.
Definite conclusions about the presence of VAPs in these films require analysis of $I$-$V$ measurements, which were not possible at the time.

In this context it is also necessary to discuss the question whether the experimental critical current is the depairing critical current of Cooper-pairs or the unpinning critical current of vortices. First of all, measured $j_\mathrm{c}$ values are only a factor of two or less smaller than the theoretically expected depairing critical current density. Furthermore, even though the Likharev criterium allows for the presence of vortices, edge barriers may prohibit the entry of single vortices due to the self-field of the applied current from the edges \cite{Stan04}. Furthermore one has to distinguish between the pinning of single vortices on the one hand and pinning of VAPs on the other. In general, for VAPs only one of the two vortices will be strongly pinned, provided that the film does have strong pinning centers. The other weakly or unpinned one is free to move under an acting Lorentz-force, if the pair should break up. This is important for the fluctuation process discussed in the following. In order to elucidate the role of vortices caused by magnetic fields a detailed study of vortex physics in such narrow superconducting strips needs to be done.
%
%Nevertheless, we assume VAPs to be existent in these structures\footnote{In this context open questions still remain with regard to the Likharev criterium, which prohibits vortices for a non-negligible temperature range between \Tc\ and \TBKT, and the role of edge barriers in narrow superconducting strips expelling vortices \cite{Stan04}.}.\marginpar{$\blacksquare$}

Coming back to the question of how VAPs might influence the count rates of random voltage pulses, we note that the count rates were measured at temperatures well below \TBKT. Free vortices present due to finite size effects will be neglected. Their average number was estimated from the areal density based on the resistance measurement and found to be negligible. Nevertheless there are a significant number of unbound vortices, although the temperature was below \TBKT. This is because of the action of the applied bias current, which in our case was very high. As mentioned above VAPs will orientate themselves with respect to the bias current near the orientation with minimum binding energy \cite{Mooij84}
\begin{equation}
U_\mathrm{b}=2\mu_\mathrm{c} + \frac{A(T)}{\varepsilon(l_j)}(l_j - 1) \label{eq.U_b}
\end{equation}
that is lower than in the undisturbed situation. Here $l_j=\ln(2.6 j_\mathrm{c}/j)$ is a current scale, $j_\mathrm{c}$ the critical current density and $j$ the applied current density, and $\varepsilon(l_j)$ is a renormalization factor close to unity. The core energy of one vortex is given by $\mu_\mathrm{c}$ and the temperature dependent energy scale $A(T)=(\pi^2\hbar/2e^2 R_\mathrm{s})\Delta(T)\tanh[\Delta(T)/\kB T]$ has to be calculated from device parameters, where $\Delta(T)$ is the superconducting energy gap at zero current. The unbinding process can be seen as thermal excitation requiring the energy $U_\mathrm{b}$. After the thermal unbinding vortex and antivortex are accelerated due Lorentz forces and move in opposite directions towards the strip edges until they are pinned at strong pinning sites or reach the strip edges where they annihilate. Assuming free flux-flow according to the Bardeen-Stephen model the voltage signal developing between the strip ends is comparable in height to the voltage pulses caused by the thermodynamic fluctuations of the order parameter described above. Also, the pulse duration is expected to be of the same order of about 1 ns or less, so that these two fluctuation modes are not easily discriminated.

For a thermally activated unbinding process the count rate should be proportional to $\exp(-U_\mathrm{b}/\kB T)$. The core energy $\mu_\mathrm{c}$ in Eq.\ (\ref{eq.U_b}) may be a non-negligible contribution to $U_\mathrm{b}$. Generally, it is assumed to be approximately proportional to the energy scale $A(T)$ \cite{Mooij84}, i.e. $\mu_\mathrm{c} \approx A(T)/\gamma$, $\gamma= \mathrm{const}$. Then the count rate may be expressed as
\begin{equation}
\Gamma_{\mathrm{BKT}}(I,T) = \Omega_{0,\mathrm{BKT}}\exp[\beta A(T)/\kB T] \left(\frac{I}{2.6\Ic}\right)^{\frac{1}{\varepsilon(l_j)}\frac{A(T)}{\kB T}}, \label{CRBKT}
\end{equation}
with $\Omega_{0,\mathrm{BKT}}$ the attempt rate and $\beta=[\gamma-2\varepsilon(l_j)]/[\gamma\varepsilon(l_j)]$.

\section{Experimental details}

The samples were made from a $d=5$ nm thick NbN film on a sapphire substrate. The film was prepared by dc magnetron sputtering in an Ar/N$_2$ gas mixture. The N$_2$ concentration was reduced with respect to the optimal concentration for a stoichiometric composition. Together with high sputtering rates (1.2 nm/s) and the substrate not being heated this resulted in films with a reduced N-content and a high degree of disorder. Consequently, the films had critical temperatures \Tc\ approximately 5 to 6 K, only about one third of the critical temperature in bulk NbN. Using a combination of electron-beam and photolithography the films were structured by reactive ion etching into a meander covering an area of roughly $4\times 4$ $\mu$m$^2$. The device used for the fluctuation measurements had a strip width $w=84$~nm and a total length $L=36\ \mu$m. Partial damage due to the etching process caused nonsuperconducting areas at the strip edges inferred from independent measurements to be about 4 nm thick. Also, thin layers (0.4 nm) at the top and bottom of the film are assumed to be normal conducting. Proximity effects associated with these normal conducting areas lead to a further reduction of the critical temperature to the above mentioned 4.81 K for the present sample. The GL coherence length extrapolated to zero temperature $\xiGL(0)=4.2$~nm was deduced from magnetoconductivity measurements and the London penetration depth $\lambda_\mathrm{L}(0) \gtrsim 300$ nm was estimated \cite{Oates91,Kubo84}. The electronic density of states at the Fermi energy $N_0 = 2.2 \cdot 10^{24}$ m$^{-3}$ K$^{-1}$ was deduced from Einstein's relation $N_0 = 1/(e^2 \rho D)$, with the diffusion coefficient $D=0.35$ cm$^2$ s$^{-1}$ determined from $B_\mathrm{c2}(T)$ near \Tc.

For measurements of the dark count rates we used a different cryogenic setup compared to the $R$-$T$ measurements in Sec.~\ref{sec.vap}. For this purpose the device was thermally anchored to the cold plate of a $^4$He bath cryostat and kept under vacuum conditions. The temperature was controlled via the vapor pressure of the helium bath. The bias current $I_\mathrm{b}$ was supplied by a low-noise custom-made biasbox in constant-voltage mode. An additional low-pass filter at the entrance to the cryostat and a voltage divider inside, close to the device, allowed to adjust the bias current reliably, even very close to \Ic$(T)$. Fluctuation events were registered as hf voltage signals which were passed on to a microwave amplifier chain with an effective band pass from 0.1 to 1.6 GHz and a total gain $\approx 50$ dB. To reduce parasitic noise the first stage was a cryogenic amplifier with a very low noise temperature of 6 K. Amplified voltage pulses were fed into either a digital 250 MHz bandwidth/1 GHz sampling rate oscilloscope or a 200 MHz bandwidth voltage-level counter. NbN meanders like to one used in this investigation can be utilized as single photon detectors with a quantum efficiency of $\approx 10 \%$ in the visible and nearinfrared spectral range \cite{Engel04,Engel04a}. Absorbed photons produce voltage signals very similar to the pulses caused by those fluctuations considered in this paper, in fact, they may be the limiting factor to the sensitivity of such detectors. Intrinsic to these detectors is a cutoff wavelength such that only photons with shorter wavelengths are detected. Assuming blackbody radiation from the surfaces facing the meander ($T \approx 4$ and 77 K, respectively), the number of detectable photons crossing the meander area is estimated to less than 10 photons/s. Considering a maximum detection efficiency of a few percent and the fact that the cut-off wavelength decreases with decreasing bias current \cite{SemenovEPJ,Engel04a}, the background photons can be safely neglected for all measurements. In addition to this background radiation from surfaces within the cryostat high energy radiation from the cosmic background radiation may also be present. We came to the conclusion that cosmic radiation can also be neglected on two causes: Firstly, the absorption probability in a 5~nm thick film is very low and, secondly, due to the high deposited energy, such quanta should trigger the resistive state even for much lower bias currents, which has never been observed.

\section{Results and discussion}
\begin{figure}
\begin{center}
 \includegraphics[width=0.85\textwidth]{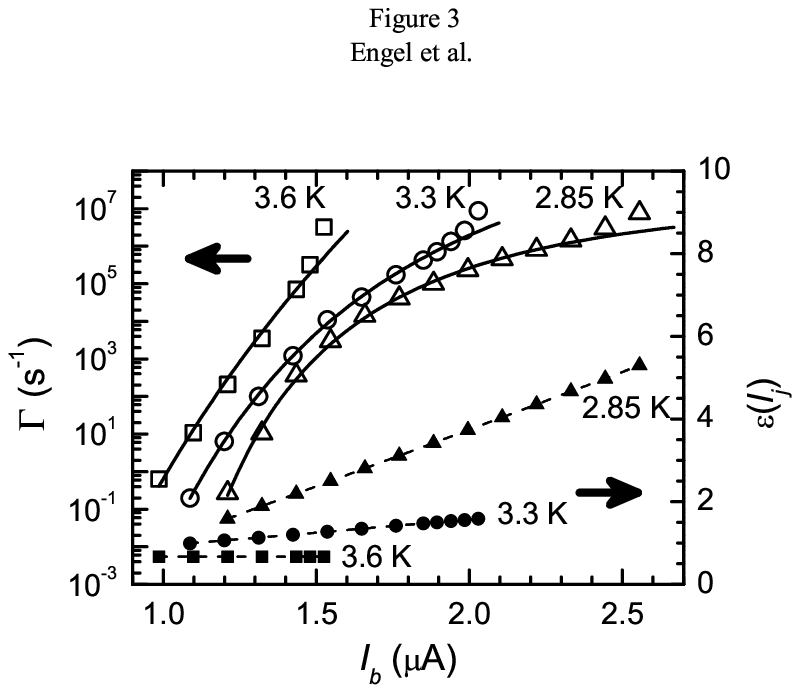}%
\end{center}
\caption{Dark count rates (open symbols, left axis, logarithmic scale) vs bias current for three different temperatures as indicated in the graph. The critical current values were $\Ic=1.55,\ 2.03$ and $2.61\ \mu$A for $T=3.6,\ 3.3$ and 2.85 K, respectively, with an accuracy of about $\pm 0.05\ \mu$A. Solid lines are least-square fits according to Eq.\ (\ref{CRBKT}), details of the fitting procedure are discussed in the text. Resulting values of $\varepsilon(l_j)$ are also plotted (right axis). \label{fig.3rates}}
\end{figure}

Voltage pulse rates were recorded as a function of the applied bias current at fixed temperatures. In Fig.\ \ref{fig.3rates} data for three different temperatures $T=3.6,\ 3.3$, and $2.85$ K are shown. At the highest temperature (squares in Fig.\ \ref{fig.3rates}) the data nearly follow an exponential dependence. However, as the temperature is reduced, the data deviate more and more from such a simple relation and open up the question as to the origin of such a temperature and current dependence. One might think of various mechanisms being able to produce voltage pulses similar to those observed in our experiments, though some of which can be ruled out. Photons of sufficient energy are certainly capable of causing such signals. However, estimates as outlined in the previous section quickly show that background radiation is many orders of magnitude too low to give any significant contribution to the observed count rates. Thermal \cite{Langer67,McCumber70} and quantum phase slips \cite{Zaikin97,Golubev01} are also expected to play a minor role, since the strip width is still considerably larger than the coherence length. Although great efforts have been made to reduce noise superimposed on the bias current contributions from current noise cannot be ruled out so easily. Especially for bias currents extremely close to \Ic\ current noise will add to the dark count rates as well as temperature fluctuations and associated fluctuations in \Ic. However, the observed temperature dependence is not easily explained by such extrinsic noise sources. Therefore, we tried to analyze the experimental data in light of the fluctuation modes discussed in the previous Sec.\ \ref{sec.pairflucs} and \ref{sec.vap}.

The solid lines in Fig.\ \ref{fig.3rates} are least-square fits of Eq.\ (\ref{CRBKT}) to the experimental data. The energy scale $A(T)$ was calculated independently for each temperature from the relevant parameters. The parameter $\beta$ in Eq.\ (\ref{CRBKT}) implicitly contains the vortex core energy $\mu_\mathrm{c}$ which may be related to the energy scale via $\mu_\mathrm{c} \approx A(T)/\gamma$ as stated in Sec.\ \ref{sec.vap}. A simple argument \cite{Mooij84} gives $\gamma \approx 8$ and if $\varepsilon(l_j)\simeq 1$ holds, one may approximate $\beta \approx 1/\varepsilon(l_j)$. Using this approximation simplifies Eq.\ (\ref{CRBKT}) considerably, leaving the attempt rate $\Omega_{0,\mathrm{BKT}}$ and $\varepsilon(l_j)$ as the only fitting parameters. We allowed for a possible current dependence of $\varepsilon(l_j)$ using a polynomial up to second order in $I_\mathrm{b}$. The resulting curves describe the data very well. Extracted $\varepsilon(l_j)$ values are also plotted in Fig.\ \ref{fig.3rates}; they show a weak or even no current dependence for the two higher temperatures that becomes much more pronounced at the lowest measured $T$. Attempt rates $\Omega_{0,\mathrm{BKT}}$ increase from $2.0\times10^5$~s$^{-1}$ over $1.2\times10^6$~s$^{-1}$ to $2.1\times10^6$~s$^{-1}$ from high to low temperature.

The general trend of increasing $\varepsilon(l_j)$ with increasing current is in line with theoretical studies \cite{Pierson95,Romano00}. A direct comparison is not possible however, because these calculations have been performed in the limit of low currents while our measurements were done in the opposite limit of very high currents. Whereas the results for $\varepsilon(l_j)$ at the two higher temperatures justify our approximation of $\beta$, it has to be called into question for the lowest $T$. The approximation results in overestimating the value of $\beta$; consequently, the derived attempt rate $\Omega_{0,\mathrm{BKT}}=2.1\times10^6$~s$^{-1}$ at $T=2.85$~K is likely to be too small.

\begin{figure}
\begin{center}
 \includegraphics[width=0.8\textwidth]{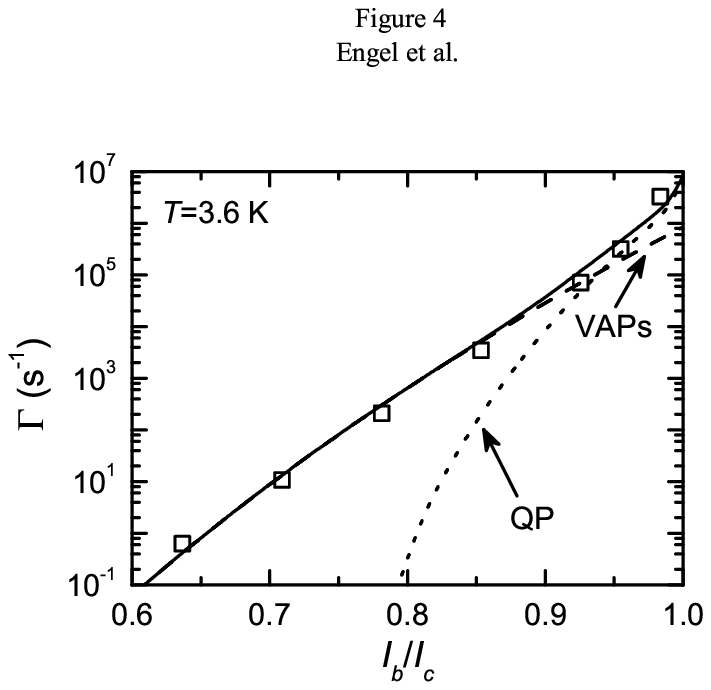}%
\end{center}
\caption{Dark count rate vs reduced bias current at $T=3.6$ K in a semi-logarithmic plot. The dashed line is the least-square fit of Fig.\
\ref{fig.3rates} and the dotted curve the dark count rate obtained for fluctuations of the number density of QP. The solid curve is the sum of these two contributions. \label{fig.BKT+cp}}
\end{figure}

As the bias current approaches the critical current the count rates for all temperatures exhibit a systematic upturn contradictory to expectations based on the unbinding of VAPs. This is when fluctuations of the QP number density may come into play. Count rates based on Eq.\ (\ref{prob1}) have to be calculated numerically and fitted to the experimental data by adjusting the attempt rate $\Omega_{0,\mathrm{cp}}$. In doing so, it soon becomes apparent that these fluctuations may only be of importance for bias currents very close to the critical current, exceeding $\approx 0.9\Ic$. As an example we show in Fig.\ \ref{fig.BKT+cp} the count rate at 3.6~K versus the reduced current $I/\Ic$. The dashed curve is the best fit according to VAP unbinding model and the dotted curve is the contribution from the QP fluctuations using an attempt rate $\Omega_{0,\mathrm{cp}} = 1.6\cdot 10^8$ s$^{-1}$. The sum of both fluctuation modes extents the excellent agreement between experiment and theoretical description up to the highest applied currents $I/\Ic \approx 0.98$. It has to be noted however, that for bias currents sufficiently close to \Ic\ current noise will
certainly add to the observed dark count rate and the observed upturn may, at least in part, be caused by this noise contribution.

\section{Conclusions}
In summary, we have described current and temperature dependent fluctuation effects in superconducting nanostructures, which were detected as voltage pulses in long meander lines. Current-assisted thermal unbinding of vortex-antivortex pairs turned out to be the most likely source of count rates over most of the current range. For high bias currents extremely close to the critical current fluctuations in the number density of QP need to be considered, as well. For the application of single-photon counters and other low-dimensional superconducting devices the detailed knowledge of the origins and control of superconducting fluctuations will be an important issue to choose the optimum operating conditions and additional investigations will be necessary. We are very grateful to D.~Golubev for stimulating discussions.

\bibliographystyle{elsart-num}
\bibliography{Literature}

\begin{thebibliography}{10}
\expandafter\ifx\csname url\endcsname\relax
  \def\url#1{\texttt{#1}}\fi
\expandafter\ifx\csname urlprefix\endcsname\relax\def\urlprefix{URL }\fi

\bibitem{Peacock96}
A.~Peacock, P.~Verhoeve, N.~Rando, A.~v. Dordrecht, B.~G. Taylor, C.~Erd,
  M.~A.~C. Perryman, R.~Venn, J.~Howlett, D.~J. Goldie, M.~Lumley, J.~Wallis,
  Single optical photon detection with a superconducting tunnel junction,
  Nature 381 (1996) 135.

\bibitem{Nakamura99}
Y.~Nakamura, Y.~A. Pashkin, J.~S. Tsai, Coherent control of macroscopic quantum
  states in a single-cooper-pair box, Nature 398 (1999) 786.

\bibitem{Skocpol75}
W.~J. Skocpol, M.~Tinkham, Fluctuations near superconducting phase transitions,
  Rep.\ Prog.\ Phys. 38 (1975) 1049.

\bibitem{Tinkham96}
M.~Tinkham, Introduction to {S}uperconductivity, 2nd Edition, {McGraw-Hill,
  Inc., New York}, 1996.

\bibitem{Knoedler82}
C.~M. Knoedler, R.~F. Voss, Voltage noise measurement of the vortex mean free
  path in superconducting granular tin films, Phys.\ Rev.\ B 26 (1982) 449.

\bibitem{Lee96}
A.~T. Lee, P.~L. Richards, S.~W. Nam, B.~Cabrera, K.~D. Irwin, A
  superconducting bolometer with strong electrothermal feedback, Appl.\ Phys.\
  Lett. 69 (1996) 1801.

\bibitem{Wilson01}
C.~M. Wilson, L.~Frunzio, D.~E. Prober, Time-resolved measurements of
  thermodynamic fluctuations of the particle number in a nondegenerate {Fermi}
  gas, Phys.\ Rev.\ Lett. 87 (2001) 067004.

\bibitem{Wilson04}
C.~M. Wilson, D.~E. Prober, Quasiparticle number fluctuations in
  superconductors, Phys.\ Rev.\ B 69 (2004) 094524.

\bibitem{Semenov04}
A.~Semenov, A.~Engel, H.-W. H{\"u}bers, K.~Il'in, M.~Siegel, Spectral response
  of an infrared superconducting quantum detector, in: B.~Culshaw, A.~G.
  Mignani, R.~Riesenberg (Eds.), Optical Sensing, Vol. 5459, Proc.\ SPIE Int.\
  Soc.\ Opt.\ Eng., 2004, p. 237.

\bibitem{Levine65}
J.~L. Levine, Dependence of superconducting energy gap on transport current by
  the method of electron tunneling, Phys.\ Rev.\ Lett. 15 (1965) 154.

\bibitem{Anthore03}
A.~Anthore, H.~Pothier, D.~Esteve, Density of states in a superconductor
  carrying a supercurrent, Phys.\ Rev.\ Lett. 90 (2003) 127001.

\bibitem{Berezinskii70}
Z.~L. Berezinskii, Zh.\ Eksp.\ Teor.\ Fiz. 59 (1970) 907.

\bibitem{Berezinskii71}
Z.~L. Berezinskii, Zh.\ Eksp.\ Teor.\ Fiz. 61 (1971) 1144.

\bibitem{Kosterlitz73}
J.~M. Kosterlitz, D.~J. Thouless, Ordering, metastability and phase transitions
  in two-dimensional systems, J.\ Phys.\ C 6 (1973) 1181.

\bibitem{Likharev79}
K.~K. Likharev, Superconducting weak links, Rev.\ Mod.\ Phys. 51 (1979) 101.

\bibitem{Aslamazov68}
L.~G. Aslamazov, A.~I. Larkin, The influence of fluctuation pairing of
  electrons on the conducitivity of normal metal, Phys.\ Lett. 26A (1968) 238.

\bibitem{Mooij84}
J.~E. Mooij, Two-dimensional transition in superconducting films and junction
  arrays, in: A.~M. Goldman, S.~A. Wolf (Eds.), Percolation, Localization, and
  Superconductivity, Plenum Press New York, 1984, p. 325.

\bibitem{Johnson98}
M.~W. Johnson, A.~M. Kadin, Electrical transport in a superconducting niobium
  nitride ultrathin film: A disordered two-dimensional {Josephson}-junction
  array, Phys.\ Rev.\ B 57 (1998) 3593.

\bibitem{Stan04}
G.~Stan, S.~B. Field, J.~M. Martinis, Critical field for complete vortex
  expulsion from narrow superconducting strips, Phys.\ Rev.\ Lett. 92 (2004)
  097003.

\bibitem{Oates91}
D.~E. Oates, A.~C. Anderson, C.~C. Chin, J.~S. Derov, G.~Dresselhaus, M.~S.
  Dresselhaus, Surface-impedance measurements of superconducting {NbN} films,
  Phys.\ Rev.\ B 43 (1991) 7655.

\bibitem{Kubo84}
S.~Kubo, M.~Asahi, M.~Hikita, M.~Igarashi, Magnetic penetration depths in
  superconducting {NbN} films prepared by reactive dc magnetron sputtering,
  Appl.\ Phys.\ Lett. 44 (1984) 258.

\bibitem{Engel04}
A.~Engel, A.~Semenov, H.-W. H{\"u}bers, K.~Ilin, M.~Siegel, Dark counts of a
  superconducting single-photon detector, Nucl.\ Instr.\ and Meth. A520 (2004)
  32.

\bibitem{Engel04a}
A.~Engel, A.~Semenov, H.-W. H{\"u}bers, K.~Il'in, M.~Siegel, Superconducting
  single-photon detector for the visible and infrared spectral range, J.\ Mod.\
  Optics 51 (2004) 1459.

\bibitem{SemenovEPJ}
A.~Semenov, A.~Engel, H.-W. H{\"u}bers, K.~Il'in, M.~Siegel, Probability of the
  resistive state formation caused by absorption of a single-photon in
  current-carrying superconducting nano-strips, submitted to Eur.\ Phys.\ J.
  Preprint available at cond-mat/0410633.

\bibitem{Langer67}
J.~S. Langer, V.~Ambegaokar, Intrinsic resistive transition in narrow
  superconducting channels, Phys.\ Rev. 164 (1967) 498.

\bibitem{McCumber70}
D.~E. McCumber, B.~I. Halperin, Time scale of intrinsic resistive fluctuations
  in thin superconducting wires, Phys.\ Rev.\ B 1 (1970) 1054.

\bibitem{Zaikin97}
A.~D. Zaikin, D.~S. Golubev, A.~van Otterlo, G.~T. Zim{\'a}nyi, Quantum phase
  slips and transport in ultrathin superconducting wires, Phys.\ Rev.\ Lett. 78
  (1997) 1552.

\bibitem{Golubev01}
D.~S. Golubev, A.~D. Zaikin, Quantum tunneling of the order parameter in
  superconducting nanowires, Phys.\ Rev.\ B 64 (2001) 014504.

\bibitem{Pierson95}
S.~W. Pierson, {$I-T$} phase diagram of vortices in layered superconductors,
  Phys.\ Rev.\ Lett. 74 (1995) 2359.

\bibitem{Romano00}
L.~Roman\`{o}, C.~Paracchini, Current effect on vortex-antivortex depairing in
  type-{II} superconductors, Phys.\ Rev.\ B 62 (2000) 5349.

\end{thebibliography}

%\begin{thebibliography}{00}

%\end{thebibliography}

\end{document}